\documentclass{article}
\usepackage{graphicx}
\usepackage{amsmath}

\newtheorem{theorem}{Theorem}
\newtheorem{acknowledgement}[theorem]{Acknowledgement}

\input{tcilatex}

\begin{document}

\author{G. Labeyrie, C. Miniatura and R.Kaiser \\
%EndAName
\textit{Laboratoire Ondes et D\'{e}sordre, FRE 2302 CNRS}\\
\textit{1361 route des Lucioles, F-06560 Valbonne}}
\title{Large Faraday rotation of resonant light in a cold atomic cloud}
\maketitle

\begin{abstract}
We experimentally studied the Faraday rotation of resonant light in an
optically-thick cloud of laser-cooled rubidium atoms. Measurements yield a
large Verdet constant in the range of 200000$%
%TCIMACRO{\UNICODE[m]{0xb0}}%
%BeginExpansion
{{}^\circ}%
%EndExpansion
$/T/mm and a maximal polarization rotation of 150$%
%TCIMACRO{\UNICODE[m]{0xb0}}%
%BeginExpansion
{{}^\circ}%
%EndExpansion
$. A complete analysis of the polarization state of the transmitted light
was necessary to account for the role of the probe laser's spectrum.

\end{abstract}

\section{Introduction}

During the past two years, we have been theoretically\ and experimentally
investigating coherent backscattering (CBS) of near-resonant light in a
sample of cold rubidium atoms \cite{Labeyrie1, Labeyrie2, Jonckeere}. CBS is
an interference effect in the multiple scattering regime of propagation
inside random media, yielding an enhancement of the backscattered light
intensity \cite{CBS}. This interference is very robust and can be destroyed
only by a few mechanisms, including Faraday rotation \cite{FaradayTh} and
dynamical effects \cite{Golub}. In the particular case of atomic scatterers,
we have shown that\ the existence of an internal Zeeman structure
significantly degrades the CBS interference \cite{Labeyrie1, Jonckeere}. The
breakdown of CBS due to the Faraday effect in classical samples has been
recently observed and studied in details \cite{Lenke}, in a situation where
the scatterers are embedded in a Faraday-active matrix. We are currently
exploring the behavior of CBS when a magnetic field is applied to the cold
atomic cloud. Since the Faraday effect is expected to be large even at weak
applied fields (of the order of $1$ $G=10^{-4}$ T), it seems relevant to
evaluate its magnitude in the particular regime of near-resonant excitation.

The Faraday effect, i.e. the rotation of polarization experienced by light
propagating inside a medium along an applied magnetic field, is a well-known
phenomenon \cite{Barron}.\ Faraday glass-based optical insulators are widely
used in laser experiments to avoid unwanted optical feedback. Due to the
presence of well-defined lines (strong resonances), the Faraday effect is
potentially strong in atomic systems, and has been extensively studied in
hot vapors \cite{AtChaud}. In addition, light can modify the atomic gas as
it propagates and induce alignment or orientation\ via optical pumping,
yielding various non-linear effects \cite{NonLinear}. However, our
experiment is quite insensitive to these effects and the scope of this paper
will only be the linear, ''standard'' Faraday rotation. Even though
laser-cooled atomic vapors appear interesting due to suppression of Doppler
broadening and collisions, few experiments on cold atoms are, to our
knowledge, reported in the litterature \cite{Yabuzaki}.

In Section 2 we expose a simple formalism to understand the main
characteristics of optical activity in an atomic system. This model, adapted
to the atomic structure of Rb, will be used in the quantitative analysis of
the experimental results. We also briefly recall in this Section the
principles of the Stokes analysis of a polarization state. The experimental
setup and procedure are described in Section 3. The results are presented in
Section 4, and compared to the model.

\section{Faraday effect and dichroism}

Let us consider a gas of two-level atoms excited by a near-resonant
monochromatic light field of frequency $\nu$. The induced electric dipole
has a component in phase with the excitation, which corresponds to the real
part of the succeptibility of the atomic medium (with a dispersive
behavior), and a component in quadrature, which corresponds to the imaginary
part of the succeptibility (absorptive behavior). The former thus relates to
the refractive index of the gas, while the later corresponds to the
absorption or scattering. At low light intensity $I\ll I_{sat}$ $\ $(where $%
I_{sat}=$1.6 mW/cm$^{2}$\ is the saturation intensity for Rubidium), the
refractive index $n$ of a dilute gas is given by :

\begin{equation}
n\left( \delta\right) \simeq1-\rho\frac{6\pi}{k^{3}}\frac{\delta/\Gamma }{%
1+4\left( \delta/\Gamma\right) ^{2}}   \label{indice}
\end{equation}
where $\rho$ is the atomic\ gas density, k $=2\pi/\lambda$ the light wave
number in vacuum, $\delta$ = $\omega-\omega_{at}$ the light detuning, and $%
\Gamma$\ the natural width ($\Gamma/2\pi=5.9$ MHz for Rb). On the other
hand, the imaginary part of the succeptibility yields the atomic scattering
cross-section $\sigma$ : 
\begin{equation}
\sigma\left( \delta\right) =\frac{3\lambda^{2}}{2\pi}\frac{1}{1+4\left(
\delta/\Gamma\right) ^{2}}   \label{sectionEff}
\end{equation}
with the usual Lorentzian line shape. This term will result in an
attenuation exp$\left( -\rho\sigma L\right) $ of the incident light as it
propagates over a distance $L$ inside the medium ; the quantity $%
b=\rho\sigma L$ is the optical thickness of the atomic sample. We thus see
that the wave will experience both a phase shift and an attenuation as it
propagates.

Let us now consider a J $=0\rightarrow$ J$^{\text{'}}=1$ transition excited
by a linearly polarized light field. A magnetic field \textbf{B} is applied
along the wavector \textbf{k}, whose direction is taken as the quantization
axis. The magnetic field displaces the resonance frequencies of the excited
state Zeeman sublevels of magnetic number $m_{e}=\pm1$ by an amount $%
m_{e}\mu B$, where $\mu=1.4$ Mhz/G. The incident linearly-polarized light
decomposes as the sum of $\sigma^{+}$ and $\sigma^{-}$waves of equal
amplitudes, which couple the unshifted ground state to excited state
sublevels of magnetic number $\pm1$ respectively. These waves thus propagate
in media with different refractive indices $n^{+}$ and $n^{-}$ and
scattering cross-sections $\sigma^{+}$ and $\sigma^{-}$, and experience
different phase shifts and attenuations. If, in a first step, we neglect the
absorption term, the two transmitted waves recombine in a linearly-polarized
wave, rotated by an angle $\theta=\frac{1}{2}(\varphi^{+}-\varphi^{-})=\frac{%
\pi}{\lambda}(n^{+}-n^{-})L$. This is the Faraday rotation angle. At small
magnetic field $\mu B/\Gamma\ll1$, the rotation angle is simply : $%
\theta\approx b\times\mu B/\left( \Gamma/2\pi\right) $. Thus, at small $B $,
the Faraday angle is simply the optical thickness $b$ times the Zeeman shift
expressed in units of the natural width $\Gamma$ (however, the
proportionality between $\theta$ and $b$ remains valid for arbitrary
magnetic field). It should be emphasized that, in atomic vapors, the Faraday
effect is very strong compared to that of standard Faraday materials (like
Faraday glasses), due to the high sensitivity of atomic energy levels to
magnetic field : for a density $\rho=$10$^{10}$ cm$^{-3}$, the Verdet
constant is $40%
%TCIMACRO{\UNICODE{0xb0}}%
%BeginExpansion
{{}^\circ}%
%EndExpansion
/G/mm$ ($4\times10^{5}%
%TCIMACRO{\UNICODE{0xb0}}%
%BeginExpansion
{{}^\circ}%
%EndExpansion
/T/mm$), more than four orders of magnitude above that of typical \ Faraday
glasses. However, for cold atomic gases, the linear increase of rotation
angle with magnetic field is restricted to a small field range of a few
Gauss (the Zeeman splitting must remain smaller than the natural width),
above which the Faraday effect decreases.

Of course, in the regime of near-resonant excitation we are dealing with,
one usually can not neglect absorption. The different attenuations for the $%
\sigma^{+}$ and $\sigma^{-}$ components cause a deformation of the
transmitted polarization as well as rotation. The polarization thus becomes
elliptic with an ellipticity (ratio of small to large axis) determined by
the differential absorption between $\sigma^{+}$ and $\sigma^{-}$ light.\
This effect is known as circular dichroism. The angle $\theta$ is then the
angle between the initial polarization and the large axis of the transmitted
ellipse. We will see how the Stokes analysis allows to extract $\theta$ and
the ellipticity from the measurements.

Although the simple picture developed above gives access to the main
mechanisms of optical activity in a an atomic gas, it does not\ accurately
describe the F $=3\rightarrow$F'$=4$ transition of the D2 line of Rb$^{85}$
used in this experiment. To expand the description to the case of a ground
state Zeeman structure, we will make the simplifying assumption that all the
transitions between different ground state Zeeman sublevels are
independent.\ We thus neglect optical pumping and coherences. The refractive
index for, say, $\sigma^{+}$ light, then writes as $n^{+}\left(
\delta,B\right) =\sum\limits_{m=-3}^{3}$p$_{m}c_{m}^{+2}n_{m}^{+}\left(
\delta,B\right) $, where the p$_{m}$ are the ground state sublevels
populations, the $c_{m}^{+}$ the Clebsch-Gordan coefficients for the various 
$\sigma^{+}$ transitions, and the $n_{m}^{+}\left( \delta,B\right) $ the
refractive indices for the Zeeman-shifted transitions (a transition from a
ground state of magnetic number $m_{g}$ to an excited state $m_{e}$ is
frequency-shifted by $\left( m_{e}g_{e}-m_{g}g_{g}\right) \mu B$, where $%
g_{e}=1/2$ and $g_{g}=1/3$ are the Land\'{e} factors for the F $%
=3\rightarrow $F'$=4$ transition of the D2 line of Rb$^{85}$). We can
express in the same way the total scattering cross-section for each circular
polarization. In the absence of magnetic field and assuming a uniform
population distribution among the ground state sublevels, the total
scattering cross-section on resonance is $\sigma\left( \delta=0\right)
\simeq0.43\times3\lambda^{2}/2\pi$, the 0.43 prefactor being the average of
the squared Clebsch-Gordan coefficient (or the degeneracy factor of the
transition $\frac{1}{3}\frac{\left( 2\text{F'}+1\right) }{\left( 2\text{F+1}%
\right) }$).

As discussed above, the polarization of the transmitted light can differ
from the incident linear polarization. It is thus necessary to fully
characterize the polarization state of the transmitted light. This can be
done using the Stokes formalism \cite{Born et Wolf}. Four quantities need to
be measured : the (linear) component of the transmitted light parallel to
the incident polarization ($I_{//}$), the (linear) orthogonal component ($%
I_{\bot}$), the (linear) component at 45$%
%TCIMACRO{\UNICODE{0xb0}}%
%BeginExpansion
{{}^\circ}%
%EndExpansion
$ ($I_{45%
%TCIMACRO{\UNICODE{0xb0}}%
%BeginExpansion
{{}^\circ}%
%EndExpansion
}$), and one of the two circular components ($I_{circ}$). The sum of the
first two is the total intensity $s_{0}$; the three other Stokes parameters
are : $s_{1}=I_{//}-I_{\bot}=2I_{//}-s_{0}$, $s_{2}=2I_{45%
%TCIMACRO{\UNICODE{0xb0}}%
%BeginExpansion
{{}^\circ}%
%EndExpansion
}-s_{0}$ and $s_{3}=2I_{circ}-s_{0}$. One can then compute the three
quantities which characterize any polarization state :

\begin{align}
P & =\frac{\sqrt{s_{1}^{2}+s_{2}^{2}+s_{3}^{2}}}{s_{0}}  \label{degreP} \\
\sin2\chi & =\frac{s_{3}}{s_{0}P}  \label{ellipt} \\
\tan2\theta & =\frac{s_{2}}{s_{1}}   \label{rotation}
\end{align}
Here, $P$ is the degree of polarization of the light, i.e. the ratio of the
intensities of the polarized component to the unpolarized one (a pure
polarization state yields $P=1$ while $P=0$ corresponds to totally
unpolarized light). Even though we would not \textit{a priori} expect any
unpolarized component, we will see that this analysis is indeed necessary in
our case.\ The quantity $e=\tan\left( \chi\right) $ is the ellipticity of
the polarized component ($e=\pm a/b$, where $a$ and $b$ are the small and
large axis of the ellipse respectively and the + or - sign denotes the sense
of rotation of the electric field). The Faraday angle $\theta$ is the angle
between the large axis of the ellipse and the direction of the incident
polarization.

\section{Description of experiment}

\subsection{\protect\bigskip Preparation of cold atoms}

The experimental setup, essentially the same as in our coherent
backscattering experiment, is described in detail elsewhere \cite{Labeyrie2}%
. A magneto-optical trap (MOT) is loaded from a dilute Rb$^{85}$ vapor (P $%
\sim10^{-8}$mbar) using six laser beams (diameter 2.8 cm, power 30 mW),
two-by-two counter-propagating and tuned to the red of the F$=3\rightarrow $%
F'$=4$ of the Rb$^{85}$ D2 line (wavelength $\lambda=780$ nm). The applied
magnetic field gradient is typically 10 G/cm. During the experiment, the MOT
(trapping beams, repumper, and magnetic field gradient) is continuously
turned on and off . The ''dark'' period is short enough (8 ms) so that the
cold atoms do not leave the capture volume and are recaptured during the
next ''bright'' period (duration 20 ms).

To characterize the cold atomic cloud, we measure its optical thickness as
described in the next subsection. The shape and size of the cloud is
recorded in 3D using fluorescence imaging, by illuminating the sample with a
laser beam detuned by several $\Gamma$. We use a time-of-flight technique to
measure the atom's $rms$ velocity, typically 10 cm/s. The atomic cloud
contains typically $3\times10^{9}$atoms with a quasi-gaussian spatial
distribution $\sim$ 5 mm FWHM (on average, the cloud being usually slightly
cigar-shaped), yielding a peak density of $\sim10^{10}$ cm$^{-3}$.

\subsection{Optical thickness and transmitted polarization measurements}

The laser probe used for transmission measurements lies in the horizontal
plane containing 4 of the trapping beams, at an angle of 25$%
%TCIMACRO{\UNICODE{0xb0}}%
%BeginExpansion
{{}^\circ}%
%EndExpansion
$. It is produced by a 50 mW SDL diode laser injected by a Yokogawa DBR
diode laser, whose linewidth is 2-3 MHz FWHM as estimated from the beatnote
between 2 identical diodes.\ This laser is passed through a Fabry-Perot
cavity (transmission peak FWHM 10 MHz) before being sent through the atomic
cloud. Although this filtering does not significantly reduce the linewidth
of the laser, it strongly suppresses the spectral components in the wings of
the initial lineshape which limit the accuracy of the transmission
measurement. The frequency of the probe can be scanned in a controlled way
by $\pm50$ MHz around the $3\rightarrow4$ transition of the D2 line. The
probe beam diameter is 1-2 mm, and its polarization is linear (vertical).
The power in the probe is typically 0.1 $\mu W$, yielding a saturation
parameter s $=2\times10^{-3}$. It is turned on for 2 ms (yielding a maximum
of about 80 photons exchanged per atoms), typically 5 ms after the MOT is
switched off. The transmitted beam is detected by a photodiode after a rough
spatial mode selection by two diaphragms of diameter 3 mm, distant of 1 m.
The optical thickness measurement is performed without applied magnetic
field. As we emphasized in \cite{Labeyrie2}, simply measuring the
on-resonance transmission yields a strongly biased estimate for the optical
thickness $b$, due to the off-resonant components in the probe laser's
spectrum. To overcome this problem, one solution is to scan the laser
detuning $\delta$ and record the transmission line shape. We describe this
curve as the convolution product of the transmission line for a purely
monochromatic laser T$\left( \delta\right) =\exp\left( -b\left(
\delta\right) \right) $ with the laser line shape. If the later is known,
one can extract the optical thickness from the transmission data (for
instance from the FWHM of the transmission curve). This method is quite
accurate for large values of $b$, where the width of the transmission curve
is only weakly dependent on the laser's linewidth. For small values of \ $b$%
, measuring the transmission at $\delta=0$ is more accurate but still
requires some knowledge of the probe laser's spectrum. When working with
dense atomic clouds at non-zero detuning, one should also keep in mind the
possible influence of ''lensing'' effect (focussing or deflection of the
transmitted beam), due to the spatially-inhomogeneous refractive index of
the sample. In our case, the rather large cloud size ($\sim5$ mm) and
moderate density ($\sim10^{10}$ cm$^{-3}$) yield a large focal length of
about 25 m for the cloud, and a small (but still measurable) lensing effect.

Using the measured size of the cloud and assuming a uniform population
distribution in the ground state, we can then obtain the peak atomic density
and the number of atoms in the sample. The maximal optical thickness
measured in our trap is $b=24$ (yielding a FWHM for the transmission curve $%
\Delta\delta\sim6\Gamma$).

As mentioned in Section 2, the Stokes analysis relies on four transmission
measurements. To perform the polarization measurement, we insert a
polarimeter in the path of the transmitted beam, as shown in fig.$1$. It
consists of a quarter-wave plate (only used for the circular component), a
half-wave plate and a glan prism polarizer (fixed). \ The rejection factor
of the polarizer is $\sim10^{-3}$. The four transmission signals ($I_{//}$, $%
I_{\bot}$, $I_{45%
%TCIMACRO{\UNICODE{0xb0}}%
%BeginExpansion
{{}^\circ}%
%EndExpansion
}$ and $I_{circ}$) are recorded as a function of the laser detuning. The
degree of polarization, ellipticity and rotation angle can then be computed
using expressions (3), (4), and (5).

\section{Results and discussion}

\subsection{Role of detuning}

Fig.$2$ shows a typical example of the raw signals obtained in the four
polarization channels necessary for the Stokes analysis detailed in Sec.$2$.
The transmitted intensity is recorded as a function of the detuning
(expressed in units of $\Gamma$) in the parallel (\textbf{A}), orthogonal (%
\textbf{B}), 45$%
%TCIMACRO{\UNICODE{0xb0}}%
%BeginExpansion
{{}^\circ}%
%EndExpansion
$ (\textbf{C}) and circular (\textbf{D}) polarization channels. All curves
have been scaled by the incident intensity. These data were obtained for a
sample of optical thickness $b=4.6$ and an applied magnetic field $B$ = 3G.
We see on curve (\textbf{B}) that more than 10\% of the incident light is
transferred to the orthogonal channel. On curves (\textbf{C}) and (\textbf{D}%
), the off-resonant detected intensities are close to 0.5, since the
incident linear polarization projection on each of these channels is 1/2.
The transmission curve (\textbf{D}), which corresponds to the $\sigma^{+}$
component, presents a minimum shifted towards positive detunings by the
Zeeman effect ; the position of the minimum corresponds roughly to the
splitting of the $m_{g}=+3\rightarrow m_{e}=+4$ transition (1.4 MHz/G),
which has a maximum Clebsch-Gordan coefficient of 1. Curves (\textbf{B}) and
(\textbf{C}) exhibit noticeable asymmetries, which we will discuss later.

From the data of fig.$2$, we computed the three curves $P\left(
\delta\right) $ (\textbf{A}), $e\left( \delta\right) $ (\textbf{B}) and $%
\theta\left( \delta\right) $ (\textbf{C}) characterizing the\ polarization
state of the transmitted light (fig.$3$). We note on the curve (\textbf{A)}\
of fig.3 that the degree of polarization $P$ is not always equal to unity,
and can be substantially smaller depending on the parameters. This
unexpected observation is due to the finite linewidth of our probe beam :
light components at different frequencies, initially all linearly polarized,
experience different rotations and deformations while passing through the
cloud. Because these components have different frequencies, the result of
their recombination, when integrated over a time long compared to their
beatnote time, is a loss of polarization (for example, two orthogonal
linearly-polarized waves of different frequencies and same intensity yield a
totally unpolarized light $P=0$). The result of the recombination of all the
spectral components is not straightforward to predict, since each frequency
is transmitted with a different intensity, ellipticity and rotation angle.
However, if we assume that all the spectral components are mutually
incoherent, the total intensity detected in each channel is the sum of \ the
intensities corresponding to all the spectral components. Thus, in order to
compare the experimental data with the model, we convolute the transmission
curves $I_{//}\left( \delta\right) ,$ $I_{\bot}\left( \delta\right) $, $I_{45%
%TCIMACRO{\UNICODE{0xb0}}%
%BeginExpansion
{{}^\circ}%
%EndExpansion
}\left( \delta\right) $ and $I_{circ}\left( \delta\right) $, as calculated
with the model of Sec.$2$, with the power spectrum of the probe laser. We
can then compute the curves $P\left( \delta\right) $, $e\left( \delta\right) 
$ and $\theta\left( \delta\right) $ using expressions (3), (4), and (5). We
stress that the influence of the laser's linewidth is quite strong : even
for a (lorentzian) linewidth of a tenth of the natural width, for an optical
thickness $b=5$ and a magnetic field $B=1G$, the loss of polarization is
already 32\% ($P=0.68$). This phenomenon also affects the estimates for the
ellipticity and the rotation angle, since these quantities reflect mainly
the polarization state of the dominant transmitted spectral component.
Indeed, if one again considers the example of the superposition of two waves
of different frequencies and polarizations (pure states), the results of the
Stokes analysis will vary continuously from one polarization state to the
other depending on the intensity ratio of the two waves. In the intermediate
regime of comparable intensities, the Stokes analysis will not describe
adequately any of the two polarizations. It should be noted that this
situation differs from the case where the studied light consist of a
polarized component plus a depolarized one \cite{Born et Wolf} ; in this
case, the Stokes analysis provides the ''correct'' result (that is, the
ellipticity and angle of the polarized component), even for arbitrarily
small proportion of polarized light. We experimentally tested the influence
of a polychromatic excitation by superimposing to the normal probe beam a
weaker one, obtained from the same laser but detuned by 80 MHz with an
acousto-optic modulator. The calculated degradation of \ $P\left(
\delta\right) $, $e\left( \delta\right) $ and $\theta\left( \delta\right) $
account well for the experimentally observed behavior.

Fig.$3$ (\textbf{B)} shows how the ellipticity $e$ of the polarized
component of the transmitted light varies with laser detuning. For $\delta>0 
$, it is mainly the $\sigma^{+}$ component of the incident light which is
absorbed, yielding a mostly $\sigma^{-}$ transmitted polarization (negative
ellipticity). On resonance ($\delta=0$) both components are absorbed in the
same proportion, an the transmitted polarization is linear ($e=0$). Since
the ellipticity depends on the differential absorption between circular
components, it is a direct measurement of the dichroism in the sample. The
curve on fig.$3$ (\textbf{C)} is the Faraday rotation angle computed from
expression (5). The on-resonance rotation in this case is about 40$%
%TCIMACRO{\UNICODE{0xb0}}%
%BeginExpansion
{{}^\circ}%
%EndExpansion
$. The solid lines in fig.$3$ are obtained with our model using the
convolution procedure discussed above ; the probe light spectrum is
described by the product of a lorentzian laser lineshape (FWHM\ = 3 MHz) by
a lorentzian Fabry-Perot transmission (FWHM =10 MHz). We account for the
observed asymmetry in the experimental curves by introducing a linear
variation in the populations of the ground state sublevels (i.e. a partial
orientation of the sample), with maximum variation $\pm20\%$ between extreme
magnetic numbers $m_{g}=\pm3.$ We have checked some other possible
mechanisms for this asymmetry, such as the proximity of the F$=3\rightarrow$%
F'$=3$ transition (121 MHz to the red) or optical pumping, but both effects
seem to play a small role. The fact that we were also able to invert the
asymmetry by varying the orientation of the magnetic field produced by the
compensation coils also favors the hypothesis of a partial orientation of
the medium. We actually do see some optical pumping effects, manifested by
variations of the measured transmission signals with time during the probe
pulse (duration 2 ms). The overall effect of optical pumping is to increase
the Faraday rotation as the number of exchanged photons (i.e. time)
increases, i.e. the measured rotation immediatly after turning on the probe
is smaller than after 2 ms of presence of the light (by about 6\%).

The comparison of fig.$3$ between experiment and theory shows that, despite
some discrepancies due to our rather vague knowledge of the probe lineshape
and to the simplicity of our model, the overall agreement is quite
satisfying.

\subsection{Role of magnetic field}

To determine the Verdet constant, it is necessary to measure the Faraday
angle as a function of the magnetic field. Fig.$4$ shows such curves
obtained for two different optical thicknesses : $b=0.75$ (solid circles)
and $b=9$ (open circles). For each curve, the on-resonance Faraday angle $%
\theta$ is scaled by the optical thickness of the cloud, since one expects
the rotation to be proportional to the optical thickness (thus, all
experimental data should lie on the same ''universal'' curve). The solid
line is the prediction of the model with a infinitely narrow probe laser.
Its slope around $B=0$ is about $10%
%TCIMACRO{\UNICODE{0xb0}}%
%BeginExpansion
{{}^\circ}%
%EndExpansion
/G$ , yielding a Verdet constant $V=20%
%TCIMACRO{\UNICODE{0xb0}}%
%BeginExpansion
{{}^\circ}%
%EndExpansion
/G/mm$ for an optical thickness $b=10$ and a sample diameter of 5 mm. The
dashed curve represents the small optical thickness limit when the laser
linewidth is taken into account, which lowers the slope at around $6%
%TCIMACRO{\UNICODE{0xb0}}%
%BeginExpansion
{{}^\circ}%
%EndExpansion
/G$.

Both curves exhibit the expected dispersive shape, with a linear increase of
the Faraday angle at small magnetic field values (where the Verdet constant
is defined), and then a decreasing rotation when the splitting between the $%
\sigma^{+}$ and $\sigma^{-}$ transitions becomes larger than the natural
width. The experimental curve for small optical thickness is quite close to
the model prediction (dashed line). For larger values of the optical
thickness, the measurements depart from this ideal situation due to the
finite linewidth of the laser : the curve for $b=9$ presents a smaller slope
around $B=0$ and the scaled rotation is globally reduced. As the optical
thickness is further increased, the transmitted light becomes increasingly
dominated by off-resonant components of the laser spectrum and the
information about the central (resonant) frequency is lost. This process is
further illustrated in the following subsection.

\subsection{Role of optical thickness}

In the ideal case of a monochromatic laser, one expects the Faraday rotation
to increase linearly with optical thickness. We thus recorded the rotation
at $\delta=0$ and a fixed value of $B$, as a function of the optical
thickness which was varied by detuning the trapping laser. The result of
such an experiment is shown in fig.$5$.\ The open circles correspond to an
applied magnetic field $B=2$ G, while the solid circles are for $B=8$ G. The
solid line is the expected evolution for $B=8$ G and a monochromatic laser;
the dashed line is the prediction of the same model for $B=2$ G. We see
that, for the higher value of $B$, the expected linear behavior is indeed
obtained, yielding a slope of about $8%
%TCIMACRO{\UNICODE{0xb0}}%
%BeginExpansion
{{}^\circ}%
%EndExpansion
/G$. The (absolute) rotation angle increases up to $\sim150%
%TCIMACRO{\UNICODE{0xb0}}%
%BeginExpansion
{{}^\circ}%
%EndExpansion
$. However, the evolution observed at smaller magnetic field ($B=2$ G,
circles) is quite different, where the data quickly depart from the linear
evolution\ even at low optical thickness and suddenly drop towards positive
values of the angle at large optical thickness. Qualitatively, this reflects
the fact that, at small applied field and large optical thickness, the
central (resonant) frequency of the probe laser is strongly attenuated and
can become smaller than other (off-resonant) spectral components. The
measured rotation angle then passes continuously from the angle of the
central frequency component to that of the dominant detuned component (in
the wings of the absorption line), which can be negative (see fig.$3$ (%
\textbf{C)}). The large dispersion of the data for $B=2$ G above $b\sim12$
is due to the important relative error in this low transmission regime. At
larger $B$\ field, the on-resonance transmission increases due to the Zeeman
splitting, and the central frequency component remains dominant for larger
values of the optical thickness (for instance, the total transmission at $%
b=20$ is around 0.1 for $B=8$ G, while it is only 3$\times10^{-3}$ for $B=2$
G).

Thus, the simple model of Section 2 provides us with a good description for
the various behaviors observed experimentally. A fair quantitative agreement
is obtained for moderate optical thickness or high magnetic field. The model
is also helpful to understand the important role played by the lineshape of
the probe laser in this experimental situation of optically-thick sample and
resonant light. The experimental data confirm the occurrence of large
Faraday effect inside our atomic cloud, with a Verdet constant in the range
of $20%
%TCIMACRO{\UNICODE{0xb0}}%
%BeginExpansion
{{}^\circ}%
%EndExpansion
/G/mm$ for a typical optical thickness of 10.

\section{Conclusion}

We have reported in this paper the measurement of large Faraday effect in an
optically-thick sample of cold rubidium atoms.\ Due to near-resonant
excitation, we need to take into account both Faraday rotation (differential
refractive index) and dichroism (differential absorption) to analyze the
experimental data. Using a very simple model for our F $=3\rightarrow$F'$=4$
transition, we obtain a good agreement with the experimental data. We
measure large Verdet constant of the order of $20%
%TCIMACRO{\UNICODE{0xb0}}%
%BeginExpansion
{{}^\circ}%
%EndExpansion
/G/mm$. We have shown that the finite width of the laser spectrum plays a
crucial role in the signals obtained for an optically-thick sample. A
complete analysis of the transmitted light polarization state is then
necessary to correctly interpret the data.

The determination of the Verdet constant $V$ in the cloud is an important
step in our current study of the effect of an applied magnetic field on the
coherent backscattering of light by the cold atoms. For Faraday effect to
seriously affect the CBS cone, one needs the phase difference between
time-reversed waves, accumulated on a distance of the order of the light
mean-free path $l$, to be of the order of $\pi$ (i.e. a rotation of $\pi/2$
for a linear polarization). This corresponds to a situation where $VBl$ $%
\sim1$ \cite{Lenke}. However, the main difference between the situation of
ref.\cite{Lenke} (scatterers in a Faraday-active matrix) and our atomic
cloud situation is that, in our case, the Verdet constant is determined by
the density of scatterers $\rho$ ($V$ proportional to $\rho$), which in turn
fixes the mean-free path ($l$ proportional to$1/\rho$). Thus, \textit{there
is a maximum rotation per mean-free path length scale}, which is about 13$%
%TCIMACRO{\UNICODE{0xb0}}%
%BeginExpansion
{{}^\circ}%
%EndExpansion
$ (for $B=3$ G)\ according to the curve of fig.$4$ (solid line). It seems to
us interesting to study this unusual situation. Our aim is also to
understand how the Faraday effect combines to\ the other effects due to the
atom's internal structure to determine the CBS enhancement factor in the
presence of an external magnetic field.

\begin{acknowledgement}
This research program is supported by the CNRS and the PACA\ Region. We also
thank the GDR PRIMA. The contribution of J.-C. Bernard was determinant in
the development of the experiment. We are grateful to D.\ Delande for some
very fruitful discussions.
\end{acknowledgement}

\bigskip

\textbf{Figures captions:}

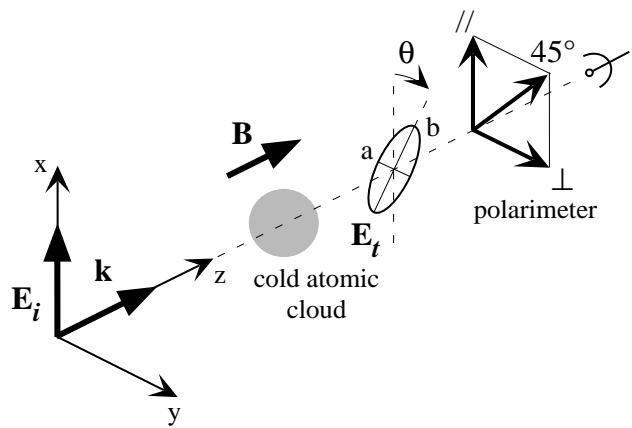
\textbf{Figure 1 :} Simplified experimental setup.

A probe laser beam of linear polarization \textbf{E}$_{\text{i}}$ and wave
vector \textbf{k} is sent through the cold atomic cloud, where a magnetic
field \textbf{B} is applied along \textbf{k}. The polarization \textbf{E}$_{%
\text{t}}$ of the transmitted probe light is deformed and rotated. A
polarimeter measures the transmitted intensities in four polarization
channels : $I_{//}$ (parallel to the incident polarization), $I_{\bot}$
(orthogonal to the incident polarization), $I_{45%
%TCIMACRO{\UNICODE{0xb0}}%
%BeginExpansion
{{}^\circ}%
%EndExpansion
}$ (at 45$%
%TCIMACRO{\UNICODE{0xb0}}%
%BeginExpansion
{{}^\circ}%
%EndExpansion
$ from the incident polarization), $I_{circ}$ (circular polarization). These
quantities allow to determine the degree of polarization $P$, the
ellipticity $e$, and the rotation angle $\theta$ of the transmitted light.

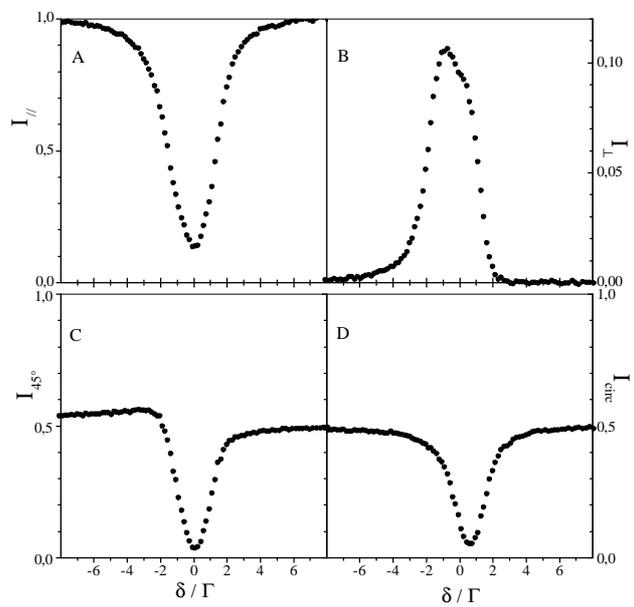
\textbf{Figure 2 : }Typical transmission curves in the four polarization
channels.

The transmission is measured as a function of the laser detuning (in units
of the natural width $\Gamma$), for a sample of optical thickness (at zero
field) $b=4.6$ and an applied magnetic field $B=3$ G. All data are scaled by
the total incident light intensity. \textbf{A} : intensity $I_{//}$ in the
linear parallel channel. \textbf{B} : intensity $I_{\bot}$ in the linear
orthogonal channel. \textbf{C} : intensity $I_{45%
%TCIMACRO{\UNICODE{0xb0}}%
%BeginExpansion
{{}^\circ}%
%EndExpansion
}$ in the linear 45$%
%TCIMACRO{\UNICODE{0xb0}}%
%BeginExpansion
{{}^\circ}%
%EndExpansion
$ channel. \textbf{D} : intensity $I_{circ}$ in the circular channel.

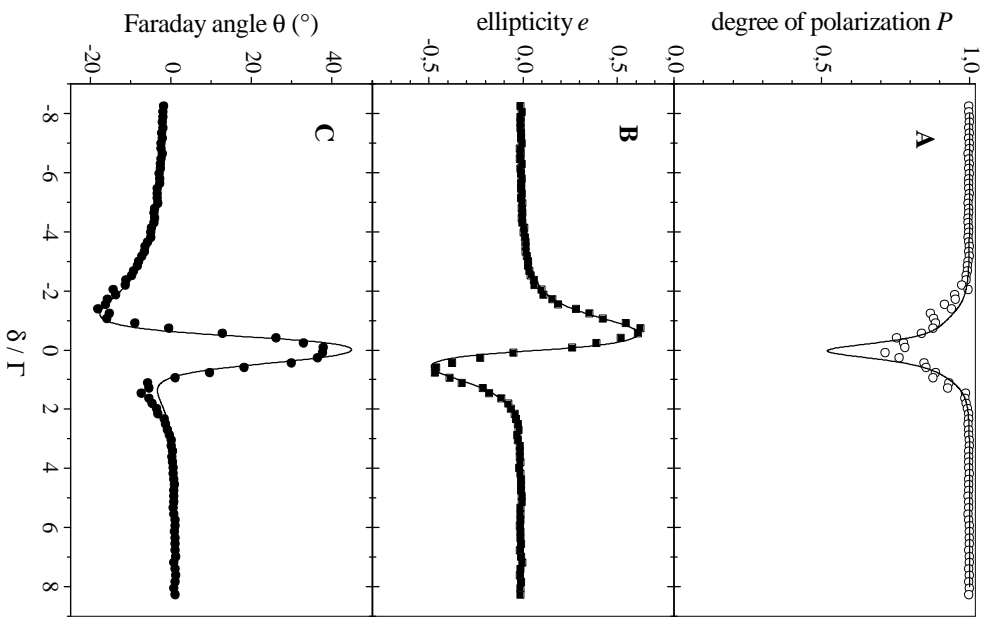
\textbf{Figure 3 : }Typical results from the Stokes polarization analysis
and comparison with model.

These curves are obtained from the data of fig.2. \textbf{A} : degree of
polarization $P$ (see expressions (3)). \textbf{B} : ellipticity $e$. 
\textbf{C} : Faraday rotation angle $\theta$. The symbols correspond to
experimental data and the solid lines to the predictions of the model
described in Sec. II. To reproduce the experimental asymmetry, the model
assumes a linear variation of the ground state populations p$_{m}$ with
magnetic number $m$, with a total variation amplitude of $40\%$ between
extreme ground state sublevels $m_{g}=\pm3$.

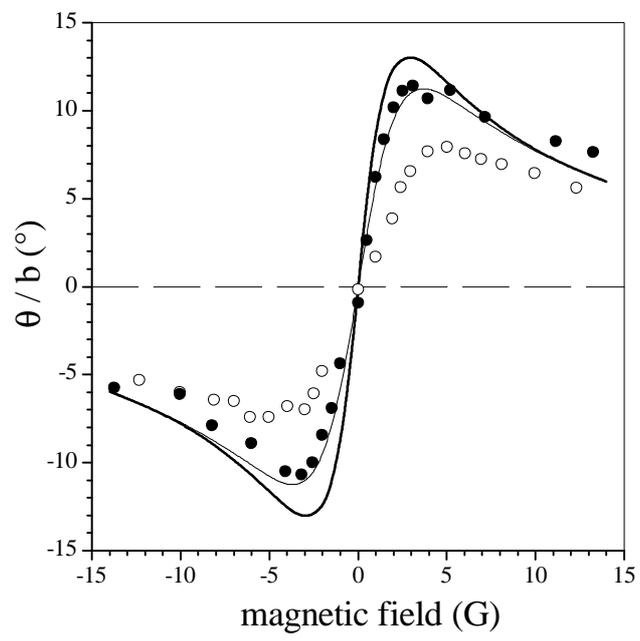
\textbf{Figure 4 :} Scaled Faraday angle as a function of the applied
magnetic field $B$.

The rotation angle $\theta$ is scaled by the optical thickness $b$ of the
sample. The symbols correspond to samples with two different optical
thicknesses : $b=0.75$ (solid circles) and $b=9$ (open circles). The solid
line is the model prediction assuming a monochromatic probe laser (and a
uniform population distribution in the ground state). The dotted line is the
small optical thickness limit of the model when taking into account the
lineshape of the probe laser.

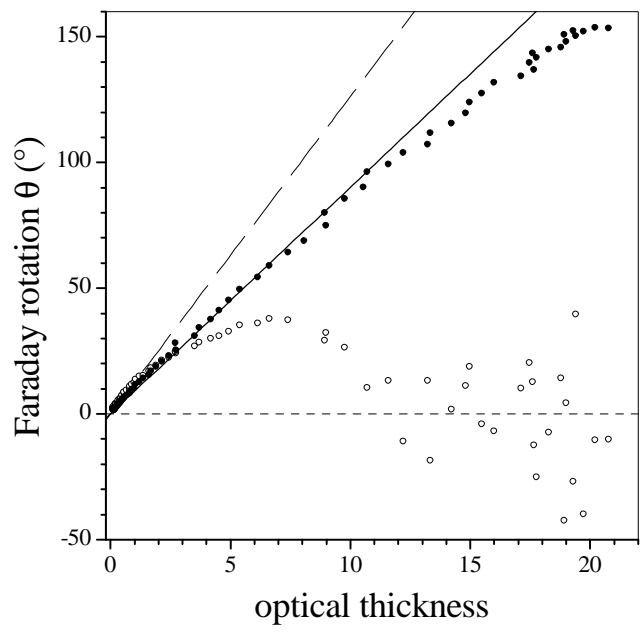
\textbf{Figure 5 :} On-resonance Faraday rotation angle $\theta$ as a
function of the optical thickness.

The symbols correspond to experiments with two different values of the
magnetic field : $B=2$ G (open circles) and $B=8$ G (solid circles). The
optical thickness of the atomic cloud is varied by scanning the detuning of
the MOT laser. The lines correspond to the predictions of the model with a
monochromatic laser, for $B=2$ G (dashed line) and $B=8$ G (solid line). For
the largest $B$ value, we observe the expected linear increase of Faraday
angle with optical thickness. The measured rotation is close to the
prediction of the ideal, monochromatic model (solid line). On the other
hand, the behavior for $B=2$ G is quite different : the rotation angle
quickly departs from the linear increase, saturates and then decreases.
Indeed, as optical thickness increases, the measured rotation becomes
increasingly affected by other spectral components of the laser, until these
off-resonant frequencies become dominant causing a sharp drop of the angle.
At large $B$ the central, resonant frequency component of the laser is
always dominant in the optical thickness range investigated, and the linear
behavior is recovered.

\end{document}